\documentclass[useAMS,usenatbib]{mn2e}
\usepackage{graphicx}
\title[Brown Dwarf Spectroscopy In Upper Scorpius]{Near Infrared Spectroscopy of Young Brown Dwarfs in Upper Scorpius}
\author[P. Dawson,  A. Scholz, T.P.Ray, D.E. Peterson, D. Rodgers-Lee, V. Geers.]{P. Dawson,$^{1}$\thanks{E-mail:
pdawson@cp.dias.ie (PD); as110@st-andrews.ac.uk (AS); tr@cp.dias.ie (TR); dpeterson@spacescience.org (DP); donna@cp.dias.ie (DR); vgeers@cp.dias.ie (VG)} 
A. Scholz,$^{2}$ T.P. Ray,$^{1}$ D.E. Peterson,$^{3}$ D. Rodgers-Lee,$^{1}$ V. Geers,$^{1}$\footnotemark[1]\\
$^{1}$School of Cosmic Physics, Dublin Institute for Advanced Studies, 31 Fitzwilliam Place, 
Dublin 2, Ireland \\$^{2}$School of Physics \& Astronomy, University of St. Andrews, North Haugh, St. Andrews, KY16 9SS, UK \\
$^{3}$Space Science Institute, 4750 Walnut Street, Suite 205, Boulder, CO 80301, USA}
\begin{document}

\date{Accepted 2013 xxxxxxx xx. Received 2013 xxxxxxx xx; in original form 2013 December xx}

\pagerange{\pageref{firstpage}--\pageref{lastpage}} \pubyear{2009}

\maketitle

\label{firstpage}

\begin{abstract}

Spectroscopic follow-up is a pre-requisite for studies
of the formation and early evolution of brown dwarfs. Here 
we present IRTF/SpeX near-infrared spectroscopy of 30 
candidate members of the young Upper Scorpius association,
selected from our previous survey work. All 24 high confidence 
members are confirmed as young very low mass objects with 
spectral types from M5 to L1, 15-20 of them are likely brown 
dwarfs. 
This high yield 
confirms that brown dwarfs in Upper Scorpius can be identified from 
photometry and proper motions alone, with negligible contamination
from field objects ($<4$\%). Out of the 6 candidates with lower confidence, 
5 might still be young very low mass members of Upper Scorpius, according 
to our spectroscopy. We demonstrate that some very low mass class
II objects exhibit radically different near infrared (0.6 - 2.5\,$\mu$m) spectra from class III objects, with strong excess 
emission increasing towards longer wavelengths and partially 
filled in features at wavelengths shorter than 1.25\,$\mu$m. 
These characteristics can obscure the contribution of the photosphere within such spectra.   
Therefore, we caution that near infrared derived spectral types 
for objects with discs may be unreliable. Furthermore, we show that 
the same characteristics can be seen to some extent in all class II and even 
a significant fraction of class III objects ($\sim 40$\%), 
indicating that some of them are still 
surrounded by traces of dust and gas. Based
on our spectra, we select a sample of 
objects with spectral types of M5 to L1, whose near-infrared
emission represents the photosphere only. We recommend the use of 
these objects as spectroscopic templates for young brown 
dwarfs in the future.

\end{abstract}


\section{Introduction}

The discovery of brown dwarfs, objects with masses below the hydrogen-burning minimum mass, at solar abundances, 
of $\sim$0.075$\,M_{\odot}$ \citep{cab00}, has challenged 
our understanding of star formation.   
Numerous theories propose several methods for the formation of substellar objects by incorporating additional physical processes 
into models of star formation.    These include; 
dynamic ejections, turbulent fragmentation, disk fragmentation, or photoerosion of cores by hot stars \citep{whi07}. 
Constraining these scenarios provides a motivation for detailed studies of young brown dwarfs in diverse environments.

The first step for all observational studies of young brown dwarfs is deep survey work
in nearby star forming regions, with the goal of identifying and characterising large samples.
These samples can then be used to constrain the mass function, the disk properties, the
binary fraction, and other diagnostics for the star formation process. An excellent target 
for this purpose is the nearby Upper Scorpius (hereafter UpSco) star-forming region \citep{pre02}, 
which hosts a large population of
brown dwarfs \citep{lodieu13}, at a presumed age of 5-10\,Myr \citep{pre02, pec12}.
UpSco is at a distance of 145$\pm$2\,pc \citep{dez99}, while \citet{pre02} note that its members are spread 
in a roughly spherical volume of approximately 20\,pc radius about this mean value.   
A recent examination of the composition of UpSco claims that it has solar metallicity \citep{mam13}.   
The region has little extinction \citep{ard00}, and measurable proper 
motion \citep{pre02}, which facilitates the identification of very low mass objects. It also has a 
disc fraction of only 23\% \citep{dsr13}.   This means that the flux received 
comes mostly from photospheres with little contamination from discs.   The young stars 
and brown dwarfs in UpSco are a dispersed population and cover 250 square 
degrees on the sky; only the advent of deep wide-field surveys has made the study 
of the entire region possible.

In our previous papers, we have identified a list of brown dwarf candidates in UpSco 
based on the Galactic Cluster Survey that was carried out as part of the UKIRT 
Infrared Deep Sky Survey \citep{dsr11, dsr13}. Our survey covered the southern part 
of the association and so it is complementary to previous work by other groups 
\citep{lodieu06, lodieu07, lodieu08, lodieu11, lodieu13, sle06, sle08}. 
Here we present spectroscopic follow-up for a large number of these candidates. 
Our aims are twofold: to verify the nature of the sources and 
to examine the spectroscopic properties of the substellar objects in this region.

\section[]{Sample}
A total of 30 objects from our survey \citep{dsr11, dsr13} were chosen for follow-up spectroscopy.   
They were chosen from a list of 96 candidate brown dwarfs identified from the UKIDSS Ninth Data Release.   
The magnitudes of the 30 are representative of those of the 96 from which they are taken.   
Their locations in UpSco are shown in Fig. 1.   
\citet{dsr13} showed that 24 of these objects have the photometric and proper motion characteristics of very low mass members of UpSco.   
As shown in the vector point diagram in Fig. 2, they form part of a population 
of objects that lie inside a 2\,$\sigma$ selection circle as calculated by \citet{dsr11} which is centred on the known proper 
motion of UpSco (-11, -25\,mas/yr \citep{deb97,pre02}) and hereinafter are referred to as ``the 2\,$\sigma$ sample''.
Of these 24 objects, 2 were first identified by \citet{lodieu06}, 13 by \citet{dsr11}, and the remaining 9 by \citet{dsr13}.   
The other 6 objects were shown in \citet{dsr11, dsr13} to have the photometric characteristics of very low mass members of UpSco.   
However, as shown in Fig. 2, they form part of a smaller population of objects that lie just outside the selection circle.

Based on mid-infrared photometry from the WISE satellite, \citet{dsr13} show that 18 of the objects from the 2\,$\sigma$ sample 
are class III objects (i.e. do not have a disc) while 6 
of them are class II objects (i.e. have a disc).   The 6 objects that lie outside the selection circle have been examined in the 
same manner as detailed in \citet{dsr13} for this work.   They exhibit the 3.4-4.6\,$\mu$m mid-infrared colours and 
weak 12 and 22\,$\mu$m signals typical of class III objects in UpSco.   
None of the 30 objects had been spectroscopically investigated before.   The 2MASS name and the position of each object are listed in Table 1.   
For the sake of simplicity and clarity, the objects are also 
numbered from 1 to 30 and 
are identified by these numbers 
throughout this work.

\begin{figure}

\includegraphics[width=0.50\textwidth]{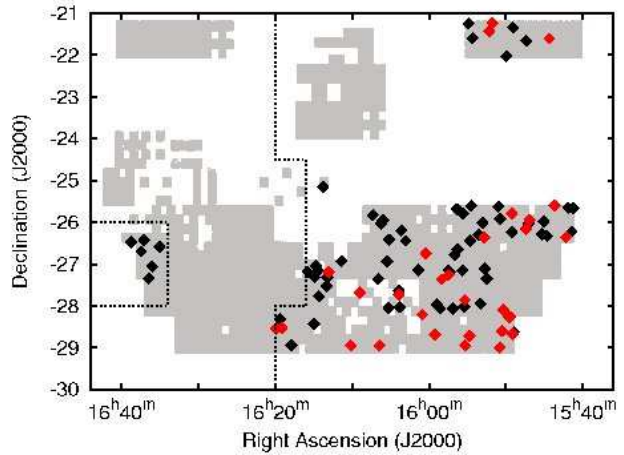}

\caption{Coverage in Z, Y, J, H and K filters of 57\,deg$^2$ in Upper Scorpius from the UKIDSS GCS.   The diamonds 
mark the position of 96 candidate brown dwarfs as identified in \citet{dsr13} from the UKIDSS Ninth Data Release.   
The 30 objects investigated in this work are shown in red.}   


\end{figure}

\begin{figure}
  \includegraphics[width=0.50\textwidth]{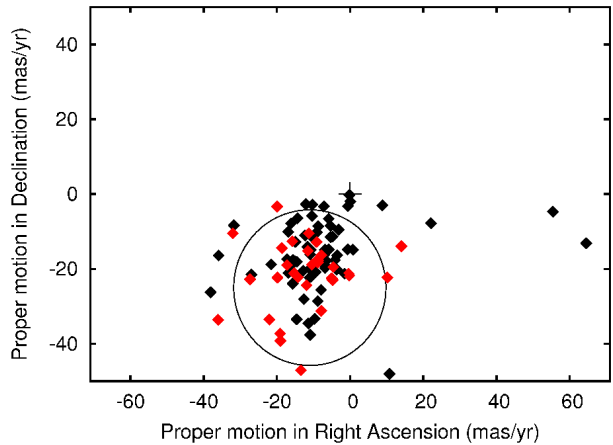}
  \caption{Vector point diagram for 96 candidate brown dwarfs in Upper Scorpius as identified in \citet{dsr13} from the UKIDSS Ninth Data Release.   
There is an obvious and identifiable cluster 
around (-11,-25), while there is no significant clustering around the origin, indicating that there is very little contamination 
from background objects in the photometrically selected sample.   The 20 candidates lying outside the 2\,$\sigma$ selection circle were classified as 
non-members by \citet{dsr13}.   The 30 objects investigated in this work are shown in red.}
\end{figure}

\section[]{Observations}
Spectra of the above targets were obtained during the nights of the 8th, 9th and 10th of June 2012 using the SpeX spectrograph \citep{rnr03} at the 
NASA Infrared Telescope Facility (IRTF) on Mauna Kea.   During the observing run, the IRTF was operated remotely from the offices of 
the Dublin Institute for Advanced Studies.   Targets were observed with SpeX in single-prism mode covering a wavelength range 
of 0.8 - 2.5\,$\mu$m.   Slits used were either 0.$''$5 or 0.$''$8 x 15$''$, depending on the seeing.   Airmass varied from 1.35 to 1.93 across 
the objects observed.   Single exposure times used were either 60, 90 or 120\,s.   Spectra were also obtained of several A0\,V stars in the 
vicinity of the 30 targets.   These spectra of standard stars were used to correct for telluric absorption in the target spectra.   
As detailed in \citet{cus04}, SpeX has an external calibration unit containing the lamps needed for flat-fielding and 
wavelength calibration.   Exposures of these lamps were taken 3 times during each observing night.      
Table 1 details the slit width used, the airmass, single exposure time and associated standard star for all 30 objects.

\begin{table*}
 \begin{minipage}{170mm}
  \caption{Summary of SpeX observations.   The first 24 objects are those in the 2\,$\sigma$ sample (see the text; Section 2).   
The last 6 objects have proper motions that caused \citet{dsr13} to classify them as non-members of UpSco (also see the text; Section 2).   
The names and positions of the objects are listed, followed by details from the observation records for each; slit width used, 
airmass, single exposure time and associated standard star.}
  \begin{tabular}{|c|c|c|c|c|c|c|c|c|c|c}
  \hline
    Object & 2MASS Name  & R.A. & Dec. & Slit Width & Airmass & Single Exposure & Standard\\
 Number & & J2000 & J2000 & (arcsec) & & Time (s) & Star \\
 \hline
1 & 2MASSJ15420830-2621138 & 15:42:08.31 & -26:21:13.8 & 0.5 & 1.73 & 60 & HD 143747\\
2 & 2MASSJ15433947-2535549 & 15:43:39.47 & -25:35:54.9 & 0.5 & 1.53 & 120 & HD 143747\\
3 & 2MASSJ15442275-2136092 & 15:44:22.75 & -21:36:09.3 & 0.5 & 1.50 & 90 & HD 143747\\
4 & 2MASSJ15465432-2556520 & 15:46:54.32 & -25:56:52.1 & 0.5 & 1.56 & 60 & HD 143747\\
5 & 2MASSJ15472572-2609185 & 15:47:25.73 & -26:09:18.5 & 0.5 & 1.82 & 60 & HD 138813\\
6 & 2MASSJ15490803-2839550 & 15:49:08.02 & -28:39:55.2 & 0.8 & 1.55 & 90 & HD 143822\\
7 & 2MASSJ15491602-2547146 & 15:49:16.02 & -25:47:14.6 & 0.5 & 1.67 & 60 & HD 138813\\
8 & 2MASSJ15492909-2815384 & 15:49:29.08 & -28:15:38.6 & 0.8 & 1.93 & 90 & HD 138813\\
9 & 2MASSJ15493660-2815141 & 15:49:36.59 & -28:15:14.3 & 0.5 & 1.92 & 90 & HD 138813\\
10 & 2MASSJ15501958-2805237 & 15:50:19.58 & -28:05:23.9 & 0.8 & 1.63 & 120 & HD 143822\\
11 & 2MASSJ15514709-2113234 & 15:51:47.09 & -21:13:23.5 & 0.5 & 1.37 & 60 & HD 136602\\
12 & 2MASSJ15521088-2125372 & 15:52:10.88 & -21:25:37.4 & 0.5 & 1.35 & 60 & HD 136602\\
13 & 2MASSJ15524857-2621453 & 15:52:48.57 & -26:21:45.4 & 0.5 & 1.49 & 60 & HD 143747\\
14 & 2MASSJ15544486-2843078 & 15:54:44.85 & -28:43:07.9 & 0.5 & 1.63 & 90 & HD 138813\\
15 & 2MASSJ15551960-2751207 & 15:55:19.59 & -27:51:21.0 & 0.5 & 1.56 & 90 & HD 143822\\
16 & 2MASSJ15572692-2715094 & 15:57:26.93 & -27:15:09.5 & 0.5 & 1.48 & 60 & HD 143747\\
17 & 2MASSJ15582376-2721435 & 15:58:23.76 & -27:21:43.7 & 0.5 & 1.50 & 90 & HD 146606\\
18 & 2MASSJ15591513-2840411 & 15:59:15.12 & -28:40:41.3 & 0.8 & 1.87 & 90 & HD 143882\\
19 & 2MASSJ16002535-2644060 & 16:00:25.35 & -26:44:06.1 & 0.5 & 1.58 & 60 & HD 143747\\
20 & 2MASSJ16005265-2812087 & 16:00:52.66 & -28:12:09.0 & 0.5 & 1.84 & 90 & HD 138813\\
21 & 2MASSJ16062870-2856580 & 16:06:28.70 & -28:56:58.2 & 0.5 & 1.68 & 90 & HD 143747\\
22 & 2MASSJ16090168-2740521 & 16:09:01.68 & -27:40:52.3 & 0.8 & 1.74 & 90 & HD 143822\\
23 & 2MASSJ16101316-2856308 & 16:10:13.15 & -28:56:31.0 & 0.8 & 1.56 & 90 & HD 143747\\
24 & 2MASSJ16195827-2832276 & 16:19:58.26 & -28:32:27.8 & 0.5 & 1.57 & 120 & HD 143822\\
\\
25 & 2MASSJ15502934-2835535 & 15:50:29.32 & -28:35:53.9 & 0.8 & 1.53 & 120 & HD 143747\\
26 & 2MASSJ15504920-2900030 & 15:50:49.19 & -29:00:03.1 & 0.8 & 1.53 & 90 & HD 146606\\
27 & 2MASSJ15551768-2856579 & 15:55:17.70 & -28:56:58.1 & 0.5 & 1.52 & 60 & HD 143822\\
28 & 2MASSJ16035601-2743335 & 16:03:56.00 & -27:43:33.6 & 0.8 & 1.48 & 90 & HD 143747\\
29 & 2MASSJ16130482-2711214 & 16:13:04.84 & -27:11:21.8 & 0.5 & 1.50 & 90 & HD 146606\\
30 & 2MASSJ16190983-2831390 & 16:19:09.82 & -28:31:39.5 & 0.5 & 1.53 & 90 & HD 143822\\

\end{tabular}
\end{minipage}
\end{table*}

\section{Data Reduction}
The raw spectra were reduced with Spextool, an IDL-based spectral reduction package designed specifically for use with data obtained 
from SpeX \citep{cus04}.   Flat-field and wavelength 
calibration exposure frames were made using the Spextool package.   The target object spectra 
and the standard star spectra were then extracted.   Telluric absorption features in the target object spectra were subsequently removed using 
the standard star spectra and the {\it xtellcor} IDL widget within the Spextool package, using the method of \citet{vac03}.    
The spectra were then normalised at the 1.25\,$\mu$m J band to facilitate comparison with each other.   The spectra 
are shown in Figs. 3 and 5 to 8.   

\begin{figure*}
\includegraphics[width=4.0cm]{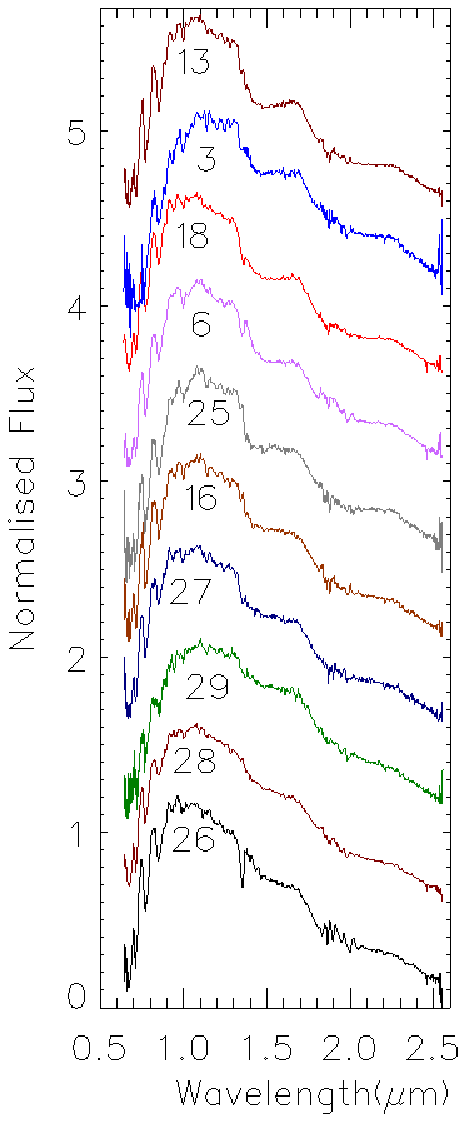} \hfill
\includegraphics[width=4.0cm]{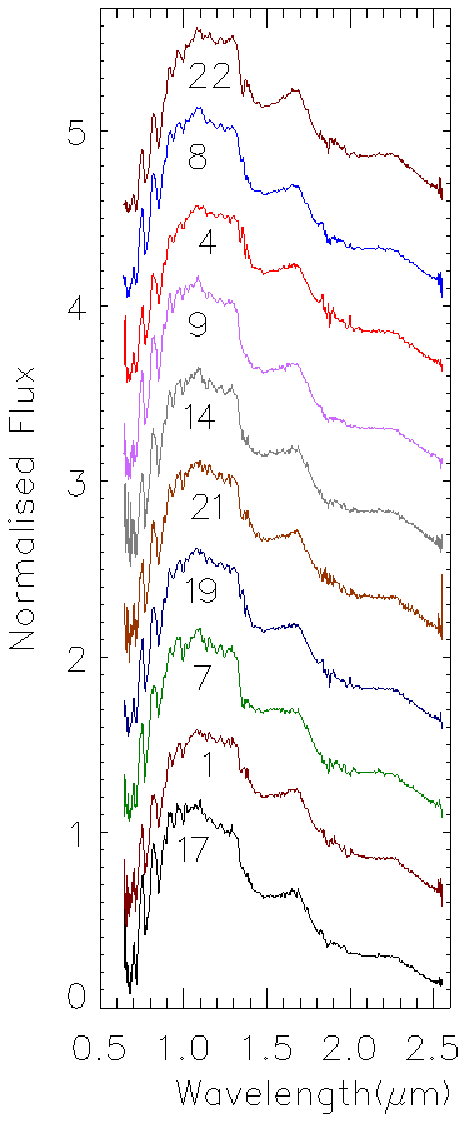} \hfill
\includegraphics[width=4.0cm]{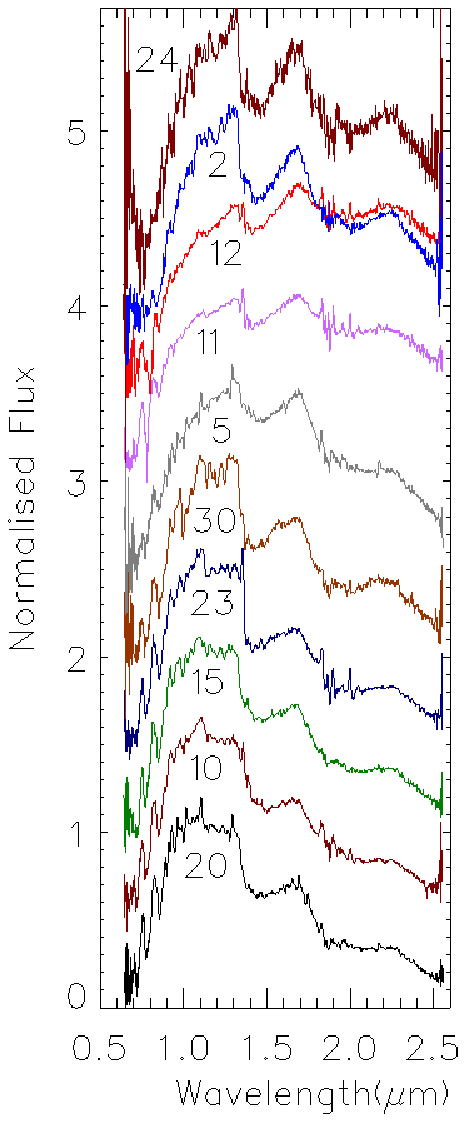} \hfill
\includegraphics[width=4.0cm]{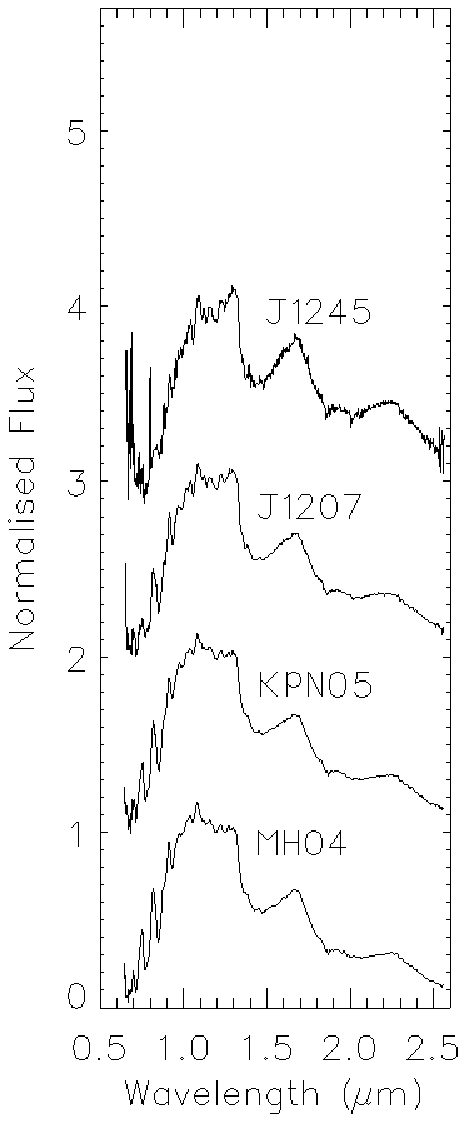}  \\
  \caption{Spectra of all 30 targets arranged in order of the prominence of the peak in the H band at 1.68\,$\mu$m.   
Also shown in the fourth panel are the spectra of 4 templates with known spectral types for comparison 
(from bottom to top: M7, M7.5, M8 and M9.5).   
Spectra are normalised at 1.25\,$\mu$m (J) and offset by 0.5.   The different profiles of the spectra of the class II 
objects (5, 11 and 12) in the third panel are quite distinct (see the text; Section 6.1).}
\end{figure*}

\section{Spectral Analysis}

The primary aim of the spectral analysis was to determine the spectral type of each object in order to confirm its nature
as a young very low mass object in UpSco.   For solar abundance brown dwarfs with 
$0.013<M<0.075\,M_{\odot}$ \citep{cab00}, we expect the spectral type to range from about M6 to L0 at the age of this region (5-10\,Myr). 
For this range of spectral types, we expect the near-infrared spectra to show a peak in the H band at 1.68\,$\mu$m and 
another in the K band at 2.25\,$\mu$m. These peaks are caused by H$_{2}$O absorption features at 1.3-1.5\,$\mu$m and 
another at 1.75-2.05\,$\mu$m and can be used as a diagnostic for objects of M-type and later. The shape of the H band peak is gravity 
sensitive, being sharply defined in young objects and much less well defined 
in evolved objects, exhibiting a plateau in the region on the blue side of 1.68\,$\mu$m \citep{kir06, pet08}.   
In addition, late-type objects exhibit a steep edge at 1.35\,$\mu$m, again caused by H$_{2}$O absorption, and CO absorption
bands at $2.3\,\mu $m. For a more detailed discussion of infrared spectra of M-L type objects, see \citet{cus05, sch09}.

\subsection{Spectral Sequence}

Inspection of the spectra showed that each object in our sample exhibited a peak in the H band at 1.68\,$\mu$m.   Its presence in 
each spectrum indicates that all the objects examined are M or L type objects.   
Later objects within this range are expected to have deeper troughs, and hence steeper slopes, 
on both the J and K side of the peak in the H band.   
The consequent prominence of the H band peak becomes correspondingly less in earlier type objects that have shallower 
troughs and slopes on either side of the peak.   
The spectra were plotted and then arranged by eye in a sequence, from earliest to latest, based on a comparison of the 
shape of each spectrum in the vicinity of the peak in the H band (first, second and third panels in Fig. 3).   
   
It is clear from Fig. 3 that object 30 in the third panel has a more rounded H band peak than the others, indicating that it is an old 
field object rather than a young substellar member of UpSco.   It is one of the 6 objects 
shown in Fig. 2 which lie outside the selection circle.   

We compared the sequence of spectra in the first, second and third panels of Fig. 3, 
with templates of young late type objects that had been given spectral types based on their optical spectrum.   
This comparison was carried out to further calibrate our visually arranged spectral sequence.
The templates were taken from the SpeX library on www.browndwarfs.org\footnote{See http://pono.ucsd.edu/$\sim$adam/browndwarfs/spexprism/}.  
The selected templates are; 
MHO4 (M7), KPNO5 (M7.5), J1207 (M8) and J1245 (M9.5), taken from \citet{mue07} and \citet{loo07} and are shown in the 
fourth panel in Fig. 3.   
Judged by their near-infrared colours, these four show little extinction ($A_V <1$\,mag, see \citet{sch12}) and so are comparable 
with the UpSco sample.  
The conclusion from this visual comparison of the templates and sample objects was that about a third of the objects were 
within half a subtype of M7 (second panel in Fig. 3), while about a third were earlier (first panel), 
and the remaining third were later than M7.5 (third panel).   

The ordering of the spectra and their visual comparison with the template objects had established that most of the objects appear to be 
young and of late M type, apart from object 30 which was determined to be an evolved field object with a type of M8.

\subsection{Spectral Indices}

In order to derive spectral types we use and compare 3 spectral definitions that have been used in the recent literature.      
The analysis of each spectrum and calculation of each index was performed using IDL routines.   

\subsubsection{The H-peak Index}
Described in \citet{sch12}, the H-peak index (HPI) is defined as the ratio of the fluxes measured in the intervals between 1.675-1.685\,$\mu$m 
and 1.495-1.505\,$\mu$m.   
The first interval in this ratio is at the position of the peak in the H band (Fig. 3).   
The second interval is located near the flux minimum on the blue side of the peak.   
The HPI therefore utilises the maximum flux range available between the bottom of the H$_{2}$O absorption band at 1.3-1.5\,$\mu$m 
and the peak in flux at 1.68\,$\mu$m.   \citet{sch12} determined an empirical relationship between the HPI and spectral types 
for the range from M7 to M9.5, noting that this relationship may 
also hold for spectral types later than M9.5, but does not apply for spectral types earlier than M7.   The correlation between the 
HPI and spectral type (SpT) is given by \citet{sch12} as

\begin{equation}
\mathrm{SpT} = -0.84 + 7.66 \times \mathrm{HPI} 
\end{equation}

In our sample, 20 of the objects are classified as M7 or later via the HPI, with L0.3 as the latest type (Table 2 and Fig. 4).  The 
remaining 10 are all given spectral 
types of between M6.1 and M6.9.   Notably, 5 of the 6 objects from outside the selection circle in Fig. 2 
are among those with spectral types earlier than M7.   The 6th object from outside the selection circle is object 30 with a 
spectral type of M8.4, but - as noted above - this is the object with the plateau near the H band peak suggesting it 
is a field object too old to be a member of UpSco.   

In summary, of the 24 objects from the 2\,$\sigma$ sample, the HPI categorises 18 of them as young very low mass objects, 
with spectral types of M7.0-L0.3, placing them all within the brown dwarf mass regime.

\subsubsection{The H$_{2}$O Index}
The H$_{2}$O index devised by \citet{all07} uses the ratio of the fluxes in the intervals between 1.550-1.560\,$\mu$m and 1.492-1.502\,$\mu$m 
to characterise spectra.   
Compared to the HPI it utilises a shorter part of the same slope on the blue side of the H band peak.   
The correlation between the H$_{2}$O index and spectral type given by \citet{all07} is 

\begin{equation}
\mathrm{H_{2}O} = 0.75 + 0.044 \times \mathrm{SpT} 
\end{equation}

\citet{all07} state that the relationship is independent of gravity and can be used to determine spectral types for field dwarfs, giants and 
young brown dwarfs over the spectral type range of M5 to L0.

Using the H$_{2}$O index, all 30 objects in our sample are classified as M5 or later with 16 of the objects classified as M6 or later, and 
with L1.1 as the latest spectral type (Table 2).   
These spectral types tend to be 1 subtype earlier than those given by the HPI (see Fig. 4), as already noted by \citet{muk12}.   
4 of the 5 objects from outside the selection circle in Fig. 2, which are classified among the earliest spectral types via the HPI, are also 
assigned the earliest spectral types (M5.2 to M5.4) via the H$_{2}$O index.   The other (object 25) is classified as an M6.4 object, while the 
field dwarf (object 30) is given a spectral type of M8.7.   

The 24 objects from the 2\,$\sigma$ sample are all given spectral types of M5.5 and later.   
Therefore the H$_{2}$O index classifies all of these 24 objects as very low mass members of UpSco ranging from the lowest stellar masses 
to the lower end of the mass range of brown dwarfs.

\subsubsection{The H$_{2}$O-K2 Index}
The H$_{2}$O-K2 index was devised by \citet{roj12} to represent the change in the overall 
shape of the spectra of M dwarfs due to water absorption in the K band from 2.07 to 2.38\,$\mu$m.   
It uses the ratio of two ratios i.e. the ratio of the fluxes between 2.070-2.090\,$\mu$m and 2.235-2.255\,$\mu$m to 
the ratio of the fluxes between 2.235-2.255\,$\mu$m and 2.360-2.380\,$\mu$m.   For objects between spectral type M0 and M9 \citet{roj12} gives the 
relationship between spectral type and H$_{2}$O-K2 index as 

\begin{equation}
\mathrm{SpT} = 24.699 - 23.788 \times \mathrm{H_{2}O-K2} 
\end{equation}

As can be seen in Table 2 and Fig. 4, the H$_{2}$O-K2 index spectral types agree with the H$_{2}$O index to within 1 subtype for 
most (23/30) of the objects.   
However, the H$_{2}$O-K2 index returns 2 markedly earlier spectral types; M3.3 and M3.7, for objects 11 and 12 respectively.   
They have HPI spectral types of M7.8 and M8.5 and H$_{2}$O index spectral types of M6.6 and M7.2.   
These are 2 of the class II objects which can be seen in Fig. 3 to have distinctly different spectral profiles to most of the other objects.   
They were shown in \citet{dsr13} to exhibit an obvious excess in their near infrared J-K colours, the result of excess emission in the K band 
from circumsubstellar discs.   So the H$_{2}$O-K2 index is measuring that part of their near infrared spectral profiles which \citet{dsr13} 
have already shown is mostly non photospheric in origin.   
Apart from these 2 objects all the objects from the 2\,$\sigma$ sample are classified with spectral types ranging 
from M4.7 to L0.8.

Of the objects that are not in the 2\,$\sigma$ sample, objects 26, 27, 28 and 29 are given spectral types varying from M2.6 to M3.7, about 2 or more 
subclasses earlier than their H$_{2}$O index spectral types.   The field dwarf, object 30, is given a spectral type of M7.5, earlier than 
its H$_{2}$O index classification of M8.7, while object 25 is assigned spectral type M6.9, similar to its H$_{2}$O index derived type 
of M6.4.   

With the exception of objects 11 and 12, the H$_{2}$O-K2 index yields spectral types of M4.7 and later for the objects 
in the 2\,$\sigma$ sample, 
which again classifies them as very low mass members of UpSco ranging from the lowest stellar masses 
down into the mass range of brown dwarfs.

\begin{figure}
  \includegraphics[width=0.50\textwidth]{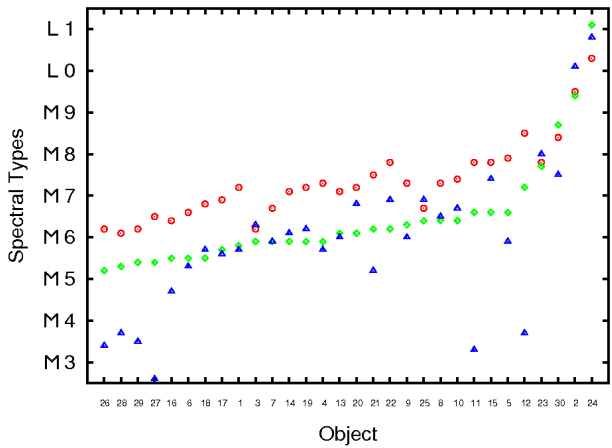}
  \caption{Spectral types of all 30 objects as derived via the H-peak (red circles), H$_{2}$O (green diamonds) and H$_{2}$O-K2 
(blue triangles) indices.   
Objects are ordered in terms of their H$_{2}$O index spectral type.
The H$_{2}$O and H-peak indices give similar spectral types for each object, the H$_{2}$O index results tending to be 1 subtype earlier.   
24 of the H$_{2}$O-K2 index spectral types agree to within 1 subtype with at least one of the other indices, 
while the remaining 6 have anomalously earlier types (see the text; Section 5.2).}
\end{figure}

\subsection{Summary Of Spectral Classifications}

We determined spectral types from comparisons with templates and with 3 different spectral indices in the H and K bands.   
The types are listed in Table 2 and shown in Fig. 4.   
The overall results are summarised as follows.

1. The 24 objects from the 2\,$\sigma$ sample have spectral types ranging from M5 to L1, 
with about 20 being of type M6 or later and they all exhibit evidence for youth.   This classifies them as likely brown dwarfs 
and young members of UpSco.

2. Of the 6 other objects, 1 of them is a field M8-M9 dwarf (and is most probably in the foreground) 
while the remaining 5 are early to mid-type (M3-M6) young M dwarfs.   
These five could still be members of UpSco, but with slightly different kinematical characteristics.

\subsection{Contamination}

\citet{dsr13} demonstrated that there was minimal contamination of the sample of 76 brown dwarf members of UpSco 
identified from the UKIDSS Ninth Data Release, based on the distribution of the proper motion vectors. 

The spectral types of the 24 objects in the 2\,$\sigma$ sample obtained via the methods outlined above 
confirm that each one is a young object with a spectral type between M5 and L1, classifying them as either brown dwarfs or
very low mass stars.   Ergo, there is no measurable contamination in this sample ($< 1/24$, i.e. $<4$\%).   
This confirms the efficacy of using photometric and proper motion data from both the UKIDSS Galactic Cluster Survey and 2MASS 
as outlined in \citet{lodieu06, lodieu07, lodieu13} and \citet{dsr11, dsr13} to reliably identify brown dwarf members of UpSco.   

\begin{figure*}
\includegraphics[width=0.49\textwidth]{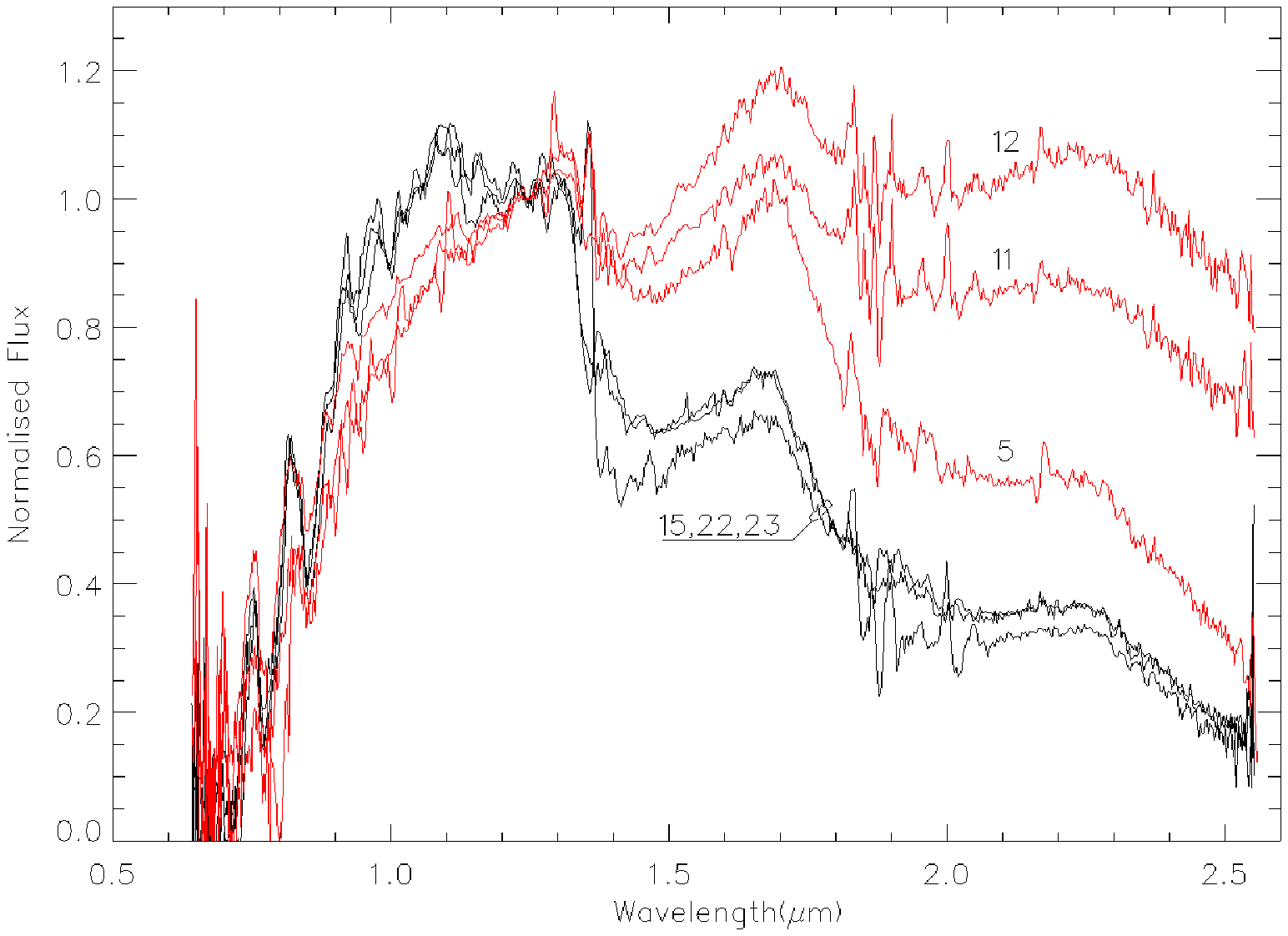}  \hfill
\includegraphics[width=0.49\textwidth]{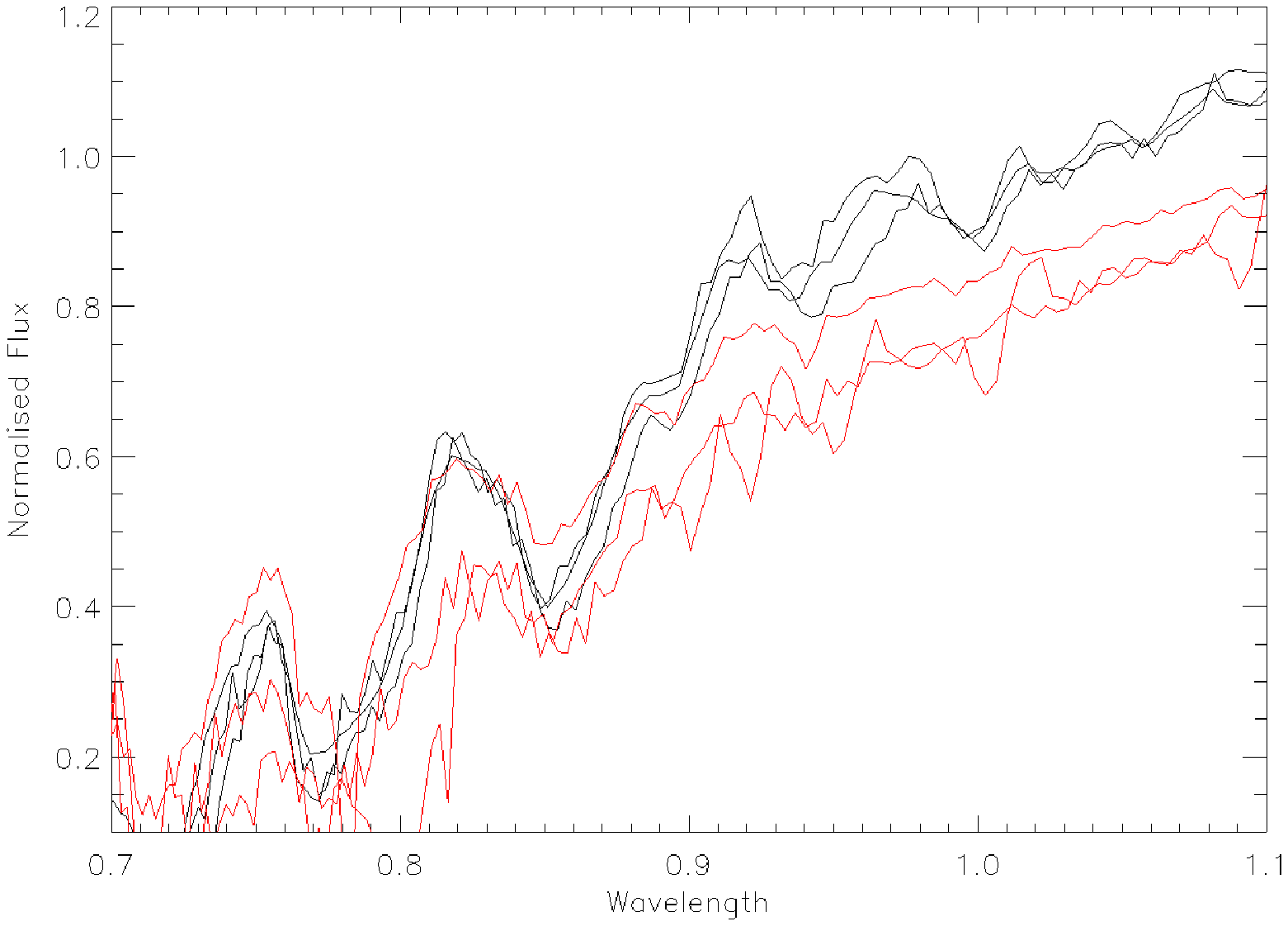}  \\

  \caption{Spectra for 3 class III objects classified as spectral type M7.8 via the H-peak index are shown in black.   
Overplotted in red are the spectra of 3 class II objects with H-peak index spectral types (reading from bottom to top) of M7.9, M7.8 and M8.5.   
All spectra are normalised at 1.25\,$\mu$m and are not offset.   
The radically different shapes of the class II spectra are readily apparent in the left panel.   
The right panel shows the shorter wavelength part of the spectra in more detail.   In this region, the class II spectra are mostly deficient 
in normalised flux and their features are weak by comparison with the class III spectra (see the text; Section 6.1).}
\end{figure*}

\section{Discussion Of Spectral Types}
Each object in our sample had spectral types assigned to it using different indices that measured specific parts of each spectrum.   
Spectra of objects with similar types were compared to each other to see how closely their overall profiles matched.

\subsection{Divergence Of Class II And Class III Spectra}

The spectra shown in Fig. 3 include 3 that display radically different profiles from all the others.   
These objects (5, 11 and 12) were shown in \citet{dsr13} to be class II objects, based on their WISE photometry in 
the W1 (3.4\,$\mu$m) and W2 (4.6\,$\mu$m) bands.   Objects 11 and 12 were also the only 2 of 27 class II objects analysed 
in \citet{dsr13} to exhibit an obvious excess in their near-infrared J-K colours.   
The differences in profile can be seen more clearly in Fig. 5 which compares these spectra with the spectra of 3 of 
the class III objects (15, 22 and 23).   
All 6 objects have similar HPI spectral types, the class III objects all being of type M7.8, while objects 5, 11 and 12 are 
of type M7.9, M7.8 and M8.5 respectively.   
The class III objects have very similar spectra.   The class II objects, on the other hand, have quite different spectra 
from the class III objects, and each other.   
However there are several features common to the class II spectra:   1; the most obvious feature is the higher normalised flux levels at 
the H and K band peaks relative to the J band,   2; the profiles of the class II spectra in the 
region on the blue side of the J band share a distinctive rounded shape, generally being deficient in normalised 
flux in this wavelength range by comparison with the class III objects,   3; features in this part of the class II spectra also tend to 
be attenuated.   
In particular, the TiO feature at 0.84\,$\mu$m tends to be much weaker among the class II objects.   
Taken together, the presence of these characteristics in a spectrum appear to be diagnostic of a class II object.   
An examination of the other class II objects in the sample (2, 4 and 10) showed that they 
exhibit these same characteristics when compared to class III objects of similar spectral type.   

The most likely origin of these spectral features is excess emission from a circumsubstellar disc.   In these objects, warm
dust from the inner disc absorbs photospheric light and emits it at longer wavelengths. This causes the observed
excess in the near-infrared bands. As seen in Fig. 5, the excess becomes stronger from the J to the K band.   
At the blue end of our spectra, the attenuation of the photospheric features is similar to the veiling 
observed in T-Tauri stars, a phenomenon which has been attributed, at least in part, to accretion processes \citep{har91, fae99}.

The presence of these characteristics raises questions about the reliability of the spectral typing of class II objects using 
near-infrared spectra.    If the flux at wavelengths longer than J contains significant contributions 
from a dusty disc, the shape of the H band or K band peak cannot be relied upon to represent the shape of the underlying photospheric spectrum.   
Fig. 5 shows that contributions from the discs of the class II objects seem to be altering the shape of the spectra 
in the region of the K band by filling in the H$_{2}$O absorption band at 1.75-2.05\,$\mu$m.   
This appears to be the reason that the H$_{2}$O-K2 index yielded much earlier spectral types for objects 11 and 12 than the other indices.     
Moreover, it is also clear from Fig. 5 that the spectra of the class II objects also have significant contributions from their discs 
in the region of the H band.   Without a clear understanding of how that part of the spectral profile is affected by contributions from 
the disc it is impossible to be confident that the spectral types derived from the H-peak and H$_{2}$O indices are any more reliable than 
the H$_{2}$O-K2 index spectral types for these 3 objects.   

Fig. 5 indicates that to obtain a reliable spectral type of a class II brown dwarf using near-infrared spectroscopy, the overall shape of the 
spectrum from 0.8 - 2.5\,$\mu$m needs to be taken into account.   If it resembles that of the class II objects in Fig. 5 any spectral 
type derived using only a small part of this wavelength range should be regarded as provisional.   
For such objects, the most reliable spectral types might still only be obtained from optical data.  
Thus, we caution against relying on 
near-infrared spectra for spectral typing of objects with known excess emission from a disc.   


\subsection{Diversity Of Class III Spectra}

The class III spectra show a lesser, but still significant diversity between objects of the same spectral type.   
In the following, we illustrate this by comparing spectra of class III objects with similar spectral types.

Fig. 6 shows the spectra of class III objects 1, 9, 13 and 19, all of which are classified between M7.1 and M7.3 using the HPI.   
The spectra have similar shapes on the blue side of the H band peak.   
However, in the H and K band, object 1 has higher normalised flux levels (by $\sim 10$\%) than the other objects and 
its features are a little shallower at the shorter wavelengths.   
The characteristics seen in the spectrum of object 1 are qualitatively similar to the appearance of the class II spectra (Section 6.1).   

The same pattern of diversity can be seen between the spectra of the class III objects 7, 14 and 19 and 
the class II object 4 in Fig. 7.   All have spectral types of M5.9 as determined via the H$_{2}$O index.   
Object 4 exhibits slight traces of the class II characteristics seen in Fig. 5.   It has higher normalised flux levels 
in the H band, while at the shorter wavelengths it exhibits a deficiency in normalized flux and attenuated features.

The spectrum of object 1 is very similar to that of object 4 in Fig. 8, where they are shown along with the class III objects 17 and 18.   
All are classified as type M5.6 or M5.7 via the H$_{2}$O-K2 index.   
The similarity of objects 1 and 4 is such that it appears that the spectrum of object 1 is not completely photospheric in 
origin but also contains small contributions from a dusty disc.   This tallies with its position in Fig 6.   
Objects 17 and 18 have lower normalised flux levels on the red side of J and display higher levels on the blue side of J, 
indicating that their spectra are more photospheric in origin.   

In summary, the differences between the spectra shown in Figs. 6, 7 and 8 are of a similar pattern to the differences between 
the class II and class III objects shown in Fig. 5, albeit at a lower level.  Compared with the cases shown in Fig. 5, the 
observed spectral diversity is small and does not affect the spectral type significantly (i.e. by more than 1 subtype).

The class II like characteristics are present in 7 of the 18 ($\sim 40$\%) class III objects in the 2\,$\sigma$ sample.   

The nature of the anomalous emission observed in some class III objects remains unclear. 
It is conceivable that the 
spectral diversity in the class III objects is related to the presence of traces of dust and gas surrounding these objects. One possibility
is that these 
objects represent an intermediate stage - a class II.5 as it were - between the objects with bona fide discs and the ``clean and clear'' 
class III sources without any evidence for excess emission.   

Alternately, the diversity among the class III spectra may be due to reddening.   
To test this, we artificially reddened the spectra of 
objects 13, 17, 18, 19 and 23, and compared them with those of the other objects.   
To redden the spectra, we used the near infrared extinction law:
\begin{equation}
\left(\frac{\mathrm{A_\lambda}}{\mathrm{A_{J}}}\right) = \left(\frac{\mathrm{\lambda}}{\mathrm{1.25\mu{m} }}\right)^{-\alpha} 
\end{equation}
with a value of 1.7 for $\alpha$, as given by \citet{mat90}, and took $A_{J}$ = 0.282$A_{\rm V}$ \citep{ral85}.   
We then adapted the method used by \citet{rnr09}, to correct for reddening in spectra obtained with SpeX, and 
reddened each spectrum using a range of values for $A_{\rm V}$ between 1 and 10.   
In each case, the least reddened spectra differed from the 
original spectrum in a manner that closely resembled the pattern of differences shown in Figs. 6, 7 and 8.

The differences that were observed are typified by the example shown in Fig. 9.   
Reddened spectra of object 19 are plotted alongside its real spectrum and compared to the spectrum of object 1.   
Both objects (which also feature in Figs. 6, 7 and 8) are classified as M7.2 using the HPI.   
The spectrum of object 1 most closely resembles the spectrum of object 19 that has been reddened using an $A_{V}$ = 1.   
An $A_{V}$ = 1 is within the limits of the little extinction that exists in the region of UpSco where 
the objects are located \citep{ard00}.
On this basis, a plausible amount of inter-stellar reddening could account for the diversity among the spectra of the class III objects.

Reddening alone is not a plausible cause of the different shape of the spectra of the class II objects shown in Fig. 5.   
Objects 5 and 11 have similar spectral types (M7.8 and M7.9 via the HPI), and yet their spectra are very different {\it to each other}, 
as well as to the spectra of the class III objects in the same diagram.   
No single reddening law could produce a match for both spectra.   
A comparison with the reddened spectra of object 23, (which has a similar HPI spectral type of M7.8), also shows that none 
of its reddened spectra are a match for the spectra of objects 5 or 11.   
Therefore, while reddening may be present in these cases, it cannot be the predominant cause of the diversity 
among these class II spectra.   
Nor, if present, can it be readily distinguished from the overwhelming contributions of the circumsubstellar discs.   

All the class II objects in this survey were shown in \citet{dsr13} to have excess flux in the mid-infrared.   
The objects with the largest mid-infrared excess were also shown to have a corresponding contiguous excess in the 
near-infrared, as shown in Figs. 4 and B1 of \citet{dsr13}.   
All of the class II objects exhibit these characteristics, although they may be present to a much lesser extent, while 
they are present in only $\sim 40$\% of the class III objects, in the 2\,$\sigma$ sample.   
Among the class III spectra that exhibit these characteristics, disentangling the relative contributions of small amounts of 
excess emission, veiling and reddening involved is not possible.   
Therefore, although near-infrared spectroscopy may be more sensitive to residual disc emission than mid-infrared photometry, it does not, 
in these cases, provide the robust certainty required for identifying circumsubstellar discs that is provided by mid-infrared photometry.   
In each case, whatever the exact cause of these effects, the underlying photospheric spectrum is being obscured and distorted.

\begin{figure*}
\includegraphics[width=0.49\textwidth]{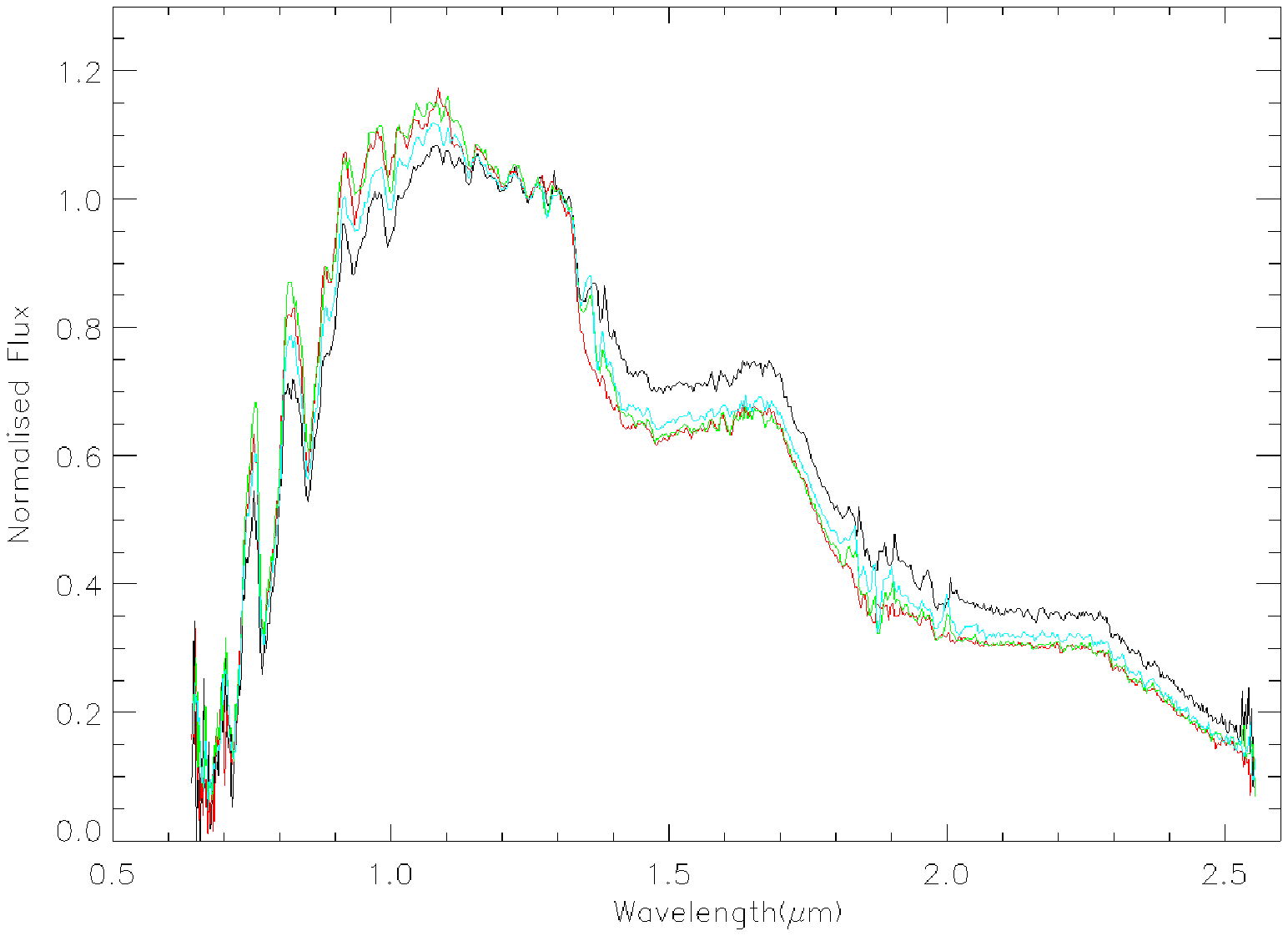}  \hfill
\includegraphics[width=0.49\textwidth]{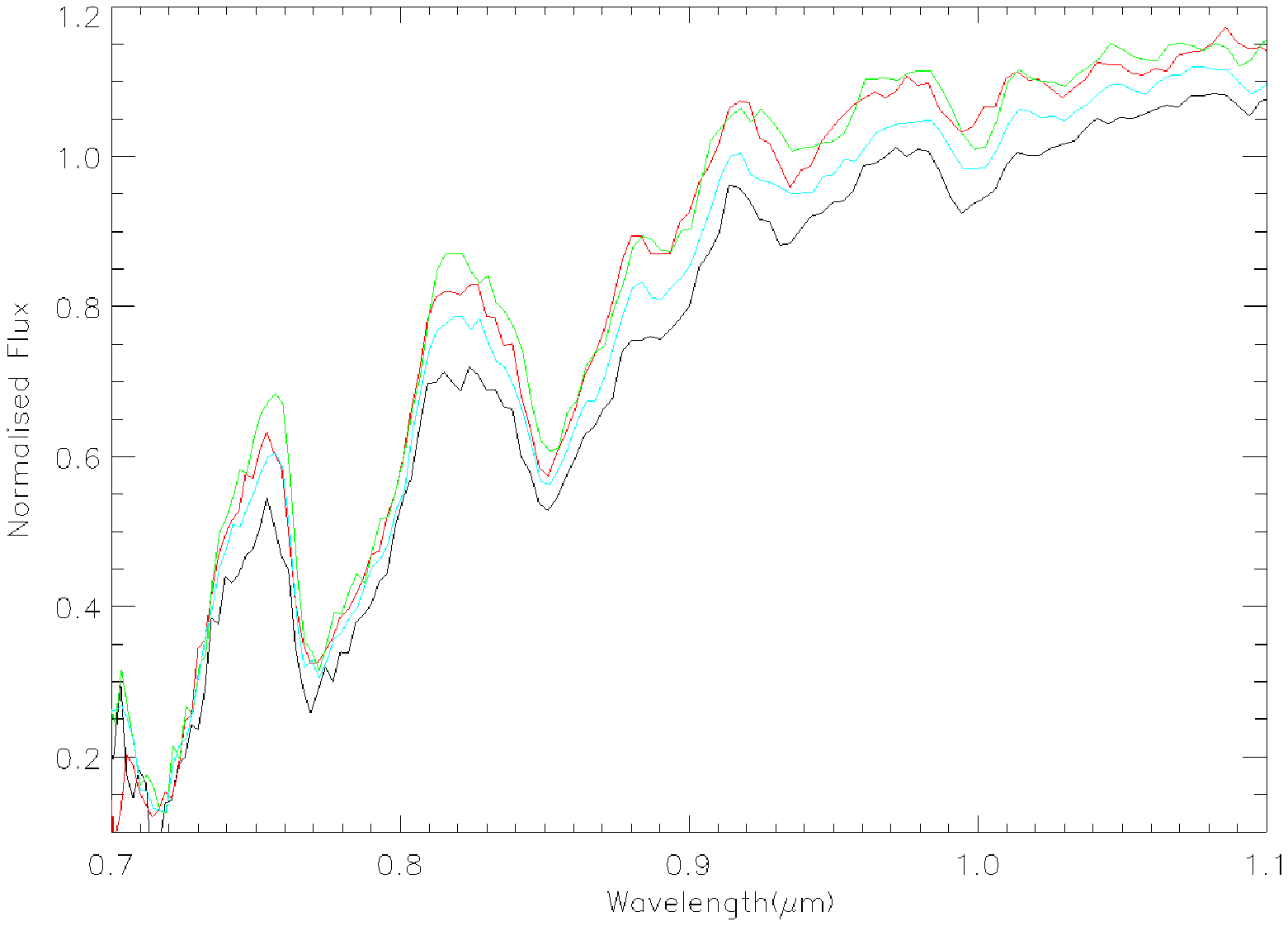}  \\
  \caption{Spectra for objects 1 (black), 9 (red), 13 (green) and 19 (blue), all classified via the H-peak index as having 
spectral types between M7.1 and M7.3.   
In the left panel the similar profiles of the spectra in the region blue of the H band peak are obvious.   
However, object 1 displays noticeably higher normalised flux levels in this region, 
and in the region blue of J most of its features are shallower than those of the other objects.   The right panel shows the latter region 
in more detail.   Object 1 has the shallowest TiO feature at 0.84\,$\mu$m.}
\end{figure*}

\begin{figure*}
\includegraphics[width=0.49\textwidth]{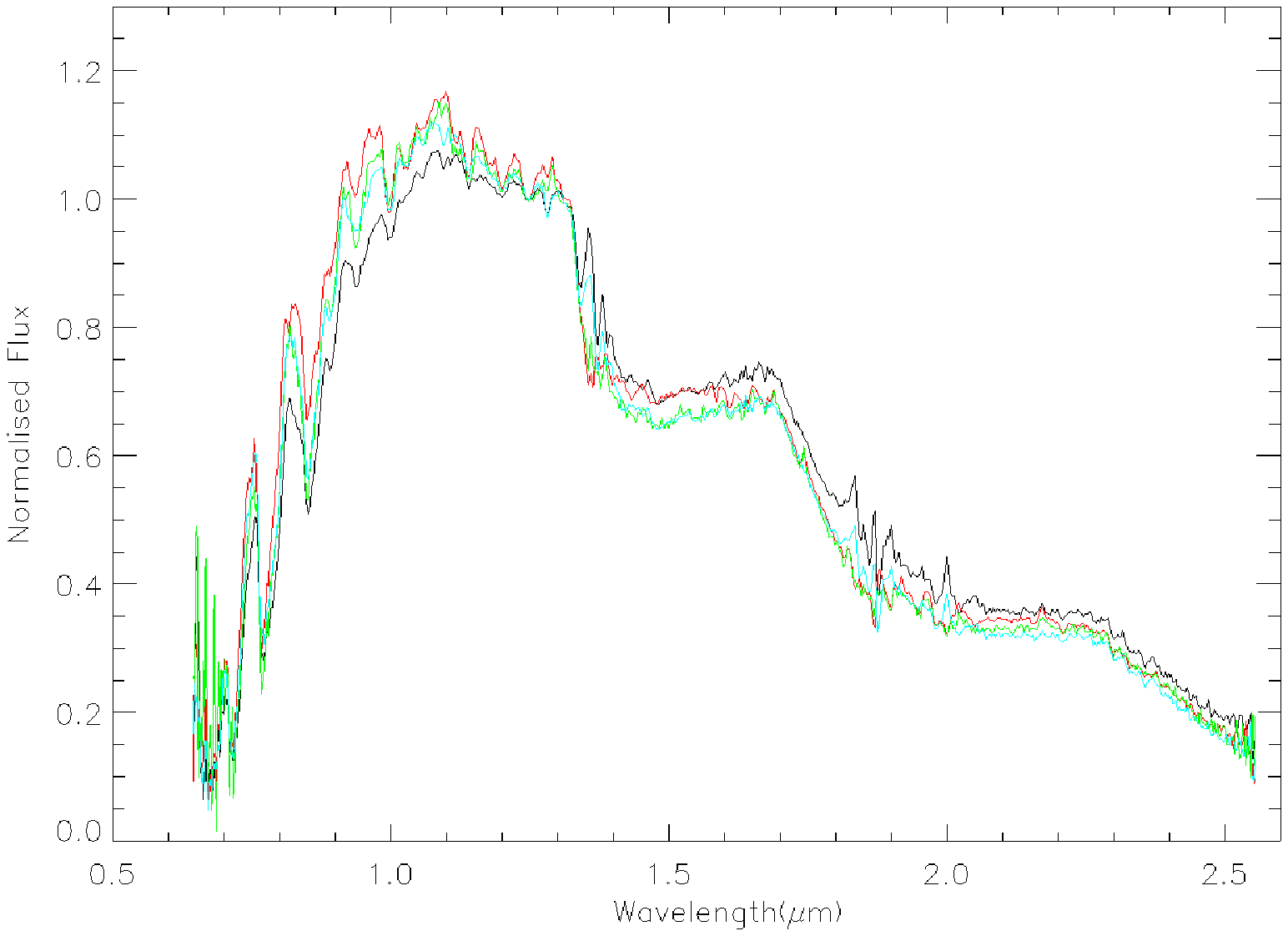}  \hfill
\includegraphics[width=0.49\textwidth]{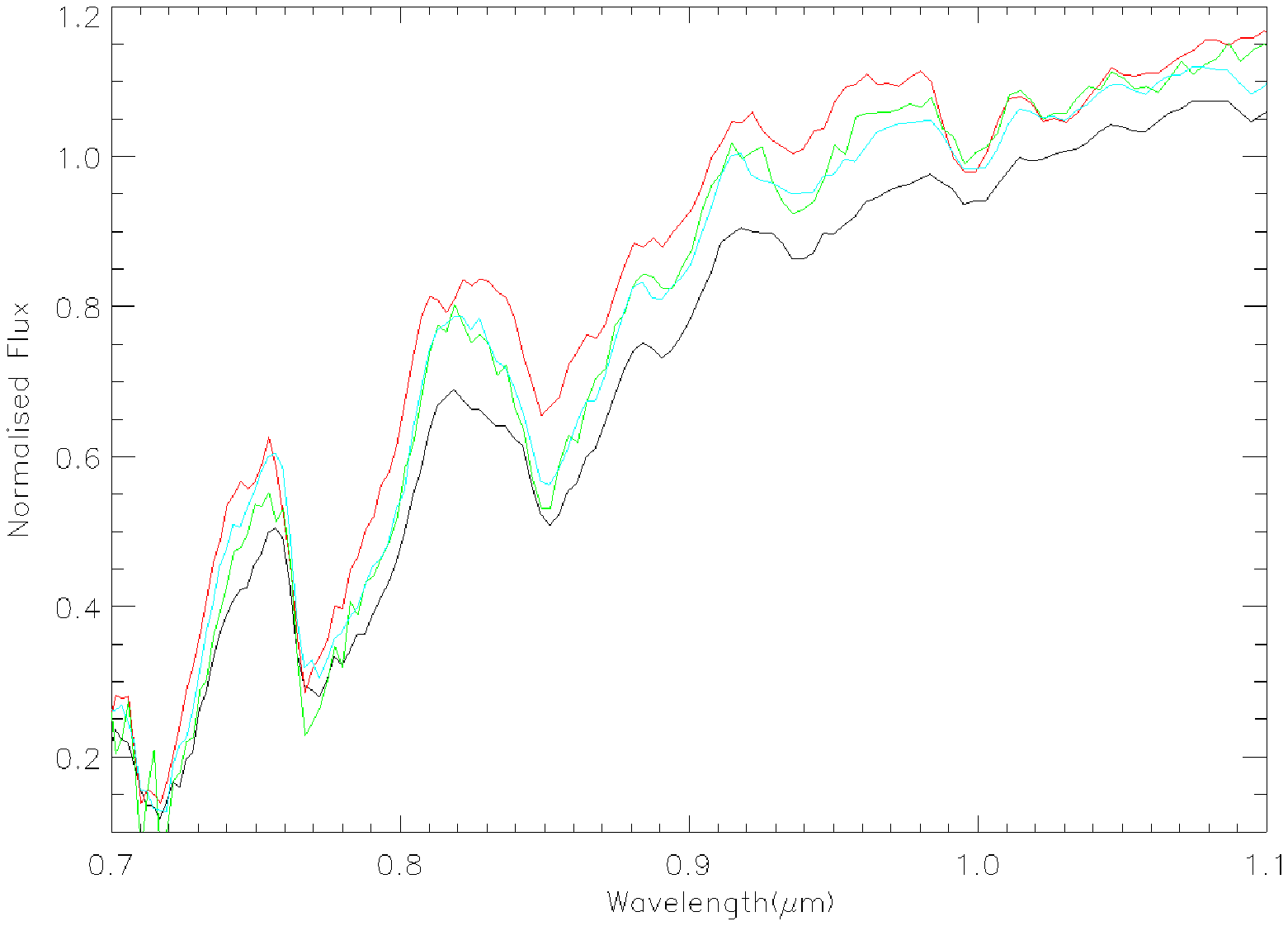}  \\
  \caption{Spectra for objects 4 (black), 7 (red), 14 (green) and 19 (blue), all classified as spectral type M5.9 via the H$_{2}$O index.   
Object 4 is a class II object and it has the highest normalised flux levels in the region red of J (left panel) along with the lowest 
levels in the region blue of J (shown in detail in the right panel) where its features are also visibly weaker by comparison with
objects 14 and 19.   These are the same traits exhibited by the class II objects shown in Fig. 5, albeit on a lesser scale.}
\end{figure*}

\begin{figure*}
\includegraphics[width=0.49\textwidth]{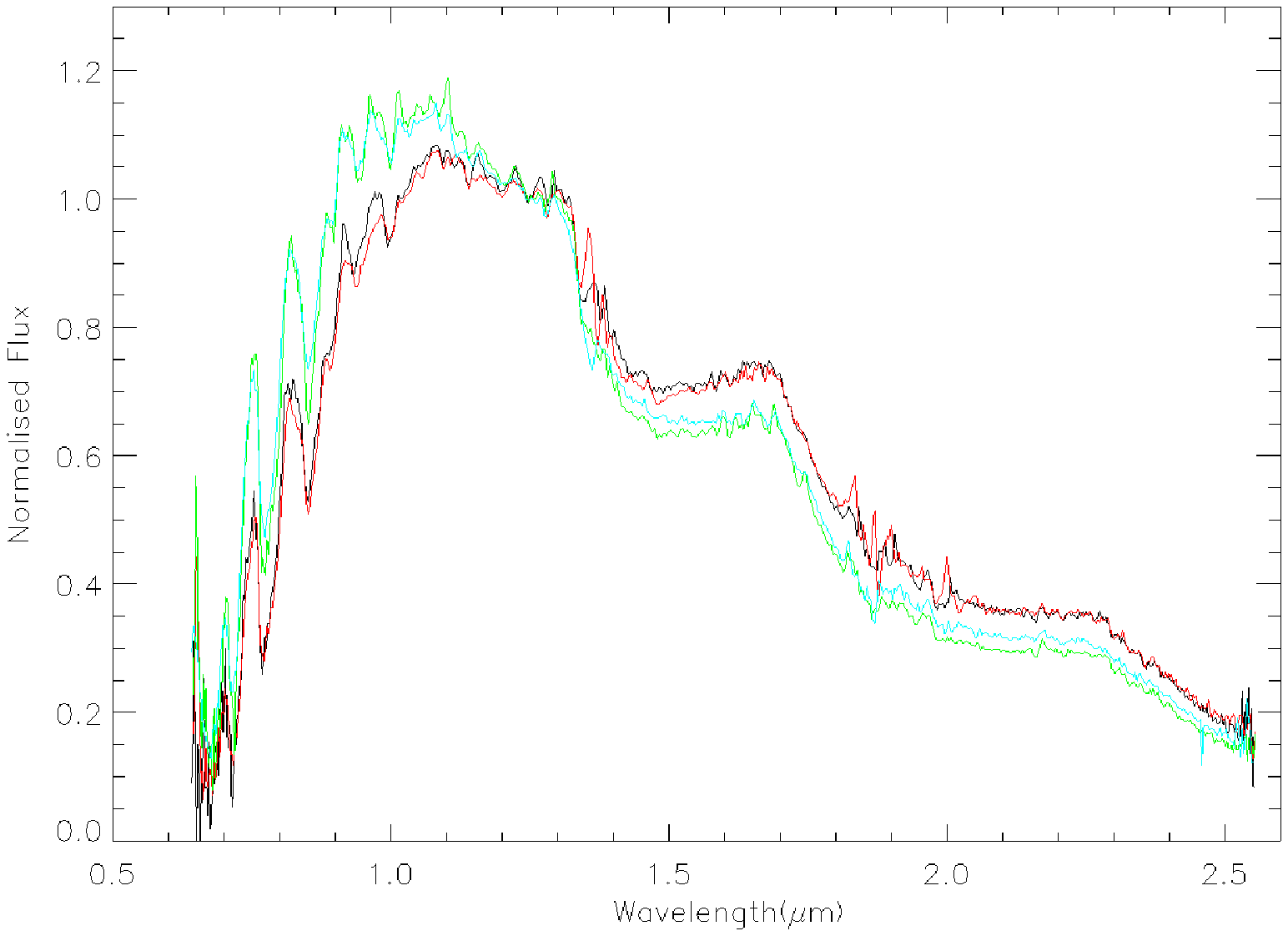}  \hfill
\includegraphics[width=0.49\textwidth]{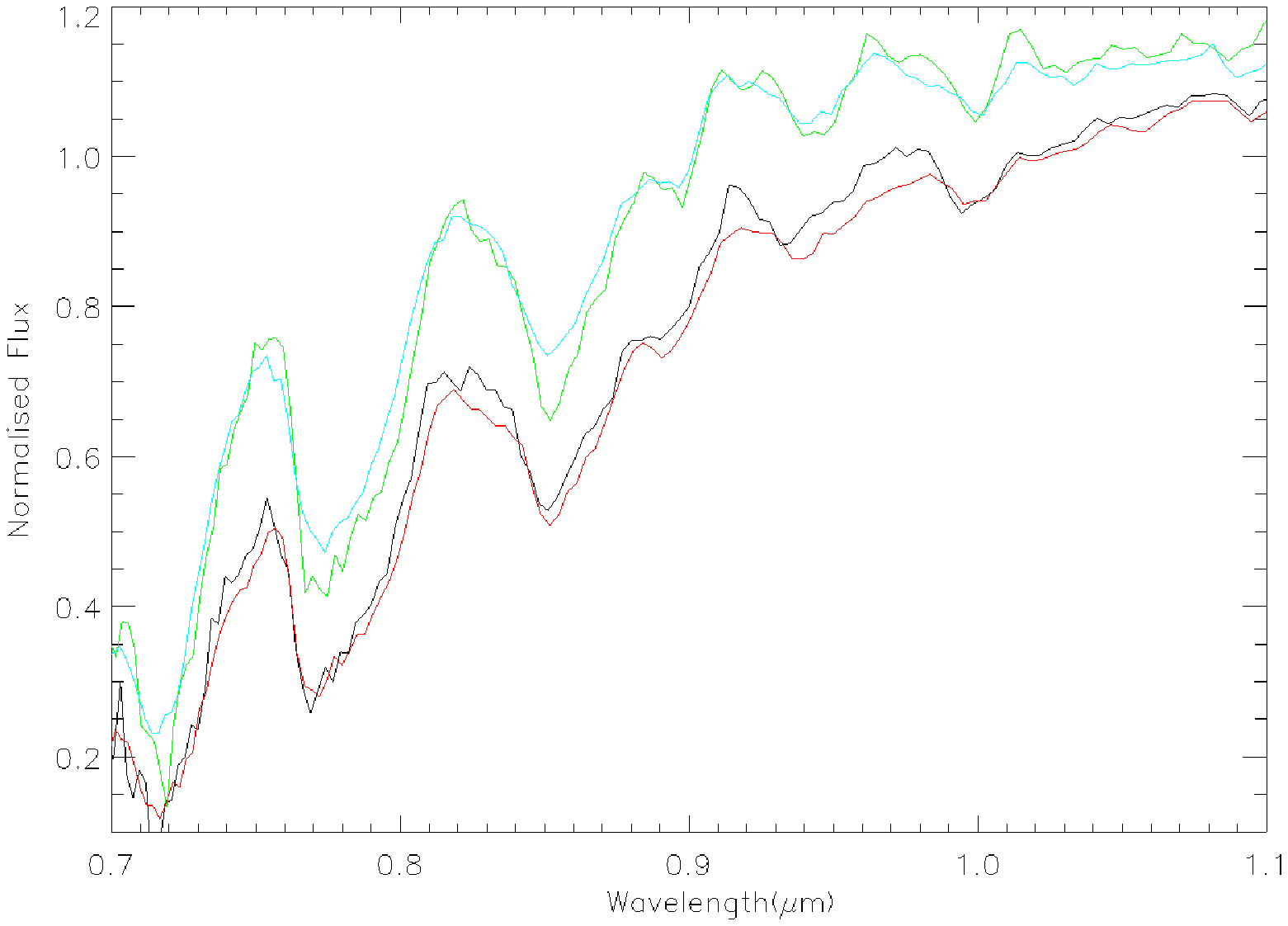}  \\  
  \caption{Spectra for objects 1 (black), 4 (red), 17 (green) and 18 (blue) classified as spectral types M5.6 and M5.7 via the H$_{2}$O-K2 index.   
In these graphs, the spectra of the class III objects 17 and 18 can be seen to be very similar to each other, and distinct 
from those of objects 1 and 4.   
The spectrum of the class III object 1 has a shape very similar to that of the class II object 4 (see the text; Section 6.2).}
\end{figure*}

\begin{figure*}
\includegraphics[width=0.49\textwidth]{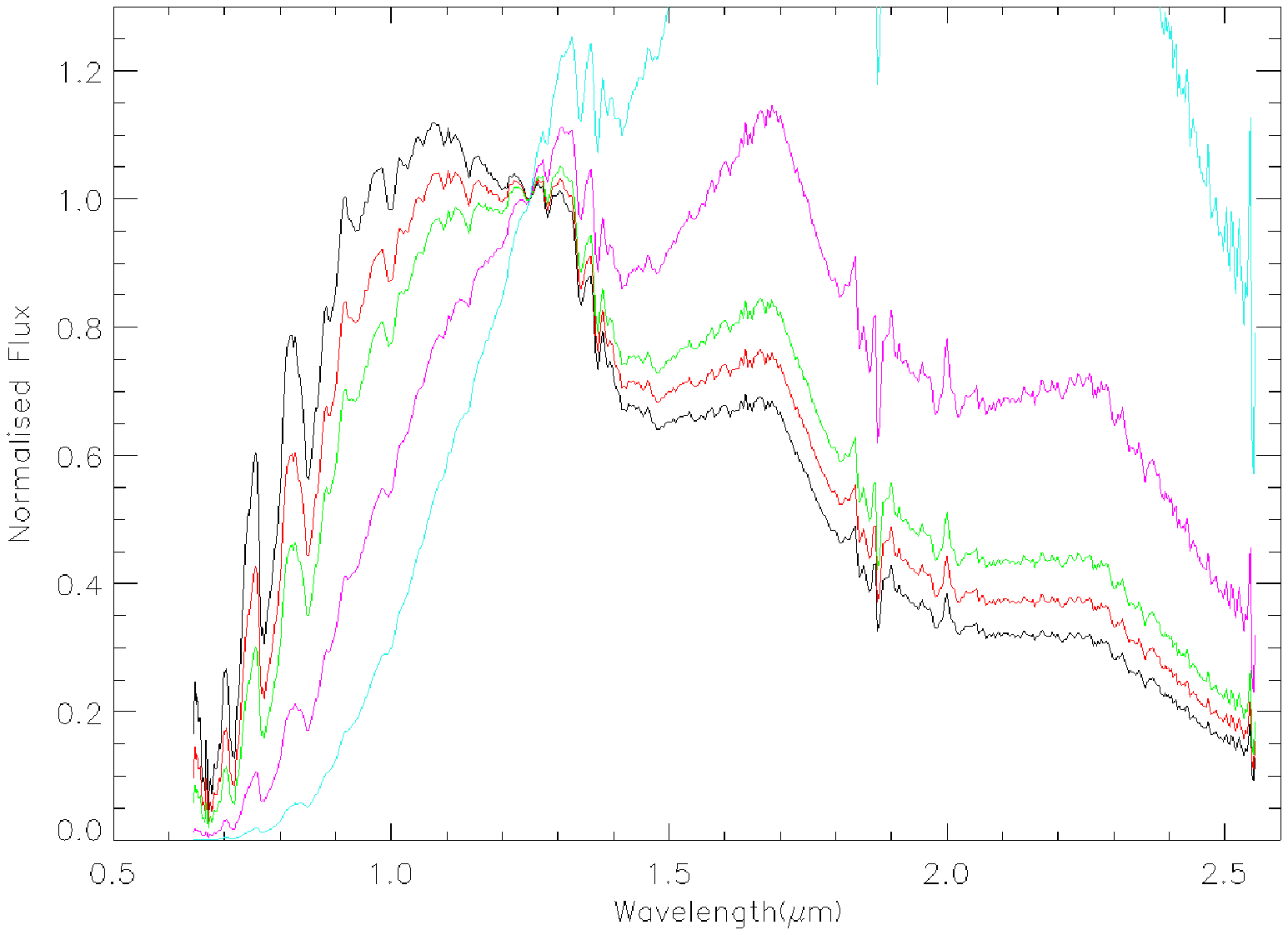} \hfill
\includegraphics[width=0.49\textwidth]{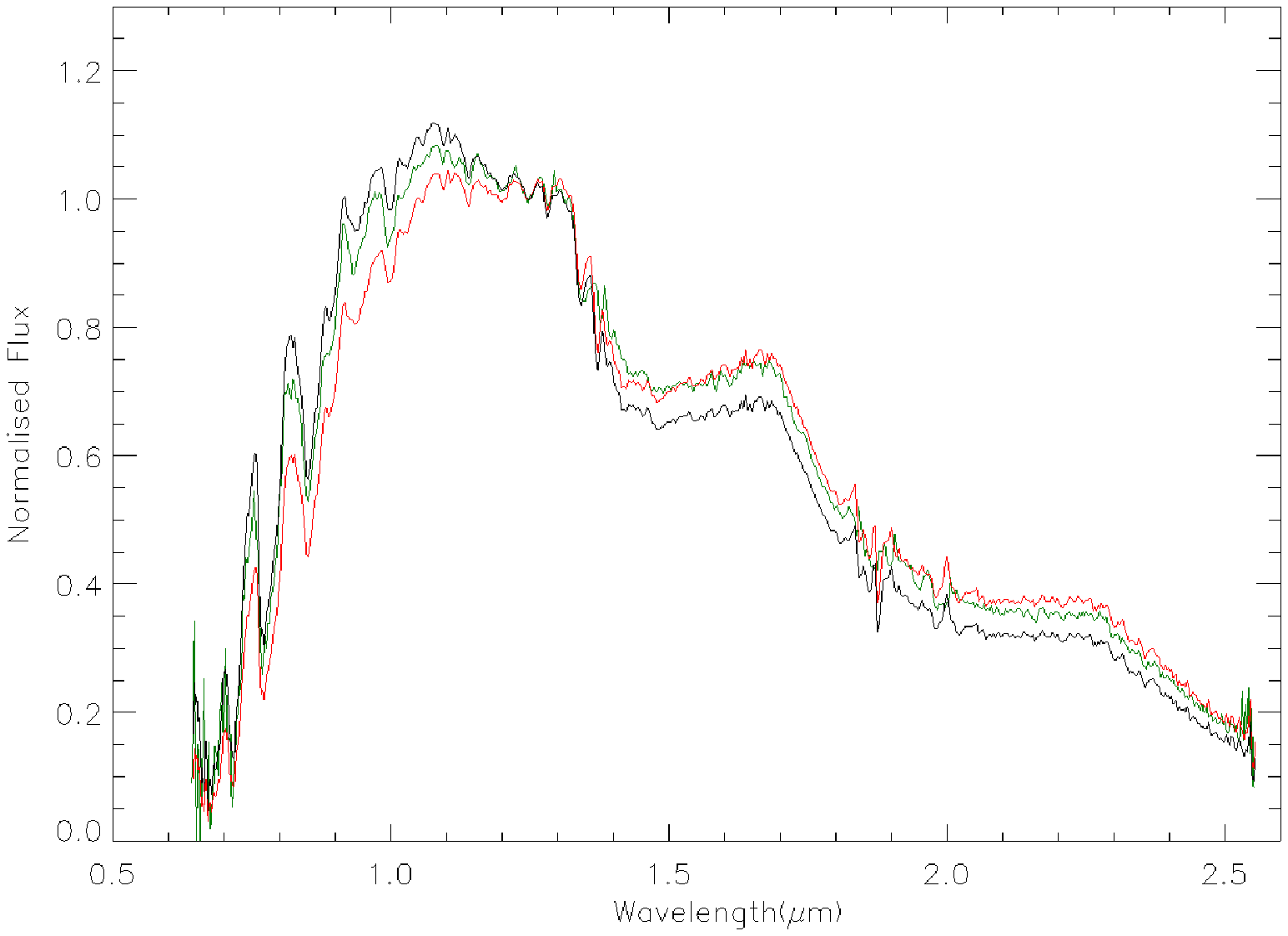}  \\  
  \caption{Spectrum of object 19, with artificially reddened spectra overplotted (left panel).   
The original spectrum (black) is the one with the lowest normalised flux in the vicinity of the H band peak.   
The 4 reddened spectra represent $A_{\rm V}$ of 1 (red), 2 (green), 5 (pink) and 10 (blue).   
The right panel shows the spectrum of object 1 (dark green), overlaid on the unreddened, and slightly reddened ($A_{\rm V}$ = 1) 
spectra of object 19 (black and red respectively).   The spectrum of object 1 is similar to the least reddened spectrum 
of object 19 (see the text; Section 6.2).}
\end{figure*}

\subsection{Templates For Spectral Typing}
There is a need for a catalogue of spectral templates of young brown dwarfs that are clean and clear of the effects of 
veiling, excess emission and reddening, 
in order to establish benchmarks that can be used to correct for these effects.   
The spectra of the 24 objects in the 2\,$\sigma$ sample were examined with the intent of producing a sequence of 
template spectra of young brown dwarfs that are free of contributions from any source other than their 
photospheres.   Our spectroscopic
survey is ideal for producing such a catalogue for two reasons: 1) UpSco is not significantly affected by extinction, 2) the disc fraction  
in UpSco (23\%) is low by comparison with other young star forming regions \citep{dsr13}, which maximises the chances of obtaining 
spectra that are purely photospheric in origin.   

In a first step,  objects 5, 11 and 12 - the class II objects from Fig. 5 - were summarily disregarded.   
Object 4 - the class II object which features in Figs. 7 and 8 - was also disregarded, along with the class II objects 2 and 10.   
Each class III object had its spectrum compared with others of a similar spectral type from each index, as in Figs. 5 to 8.   
The spectra of objects 1, 3, 6, 7, 15, 16 and 22 were found to exhibit excess emission and veiling.   
Each spectrum was also compared, in the same manner, to every other member of the 2\,$\sigma$ sample.   
irrespective of their difference in spectral types.    
Apart from the expected variety among objects of different spectral type, no other pattern of diversity or anomaly, 
other than that already discussed, was discerned.

The spectra of the remaining class III objects (Table 3) are those that, by comparison, show no evidence of excess emission, veiling 
or reddening.   
We present them as a catalogue of spectra of young brown dwarfs which, to the greatest extent discernible, are clear of contributions 
from sources other than their photospheres.   
The spectral types of the chosen objects, as determined from our near-infrared spectra, range from M5.5 to L1.1.   
Table 3 also includes a rough estimate of the effective temperature (T$_{eff}$) of each object.   The T$_{eff}$ of each object was 
estimated using the Spectral Type-T$_{eff}$ relationship given in \citet{muk14} for objects of spectral types later than M1.   
The different spectral types of each object, derived via the 3 different spectral indices, yielded a range of values of T$_{eff}$ for 
each object.   
The T$_{eff}$ listed in Table 3 is the median of these values, rounded to the nearest 50K.   
The typical uncertainty in each estimated T$_{eff}$ is 150-200K.   
The catalogue of these spectra can be accessed at: http://browndwarfs.org/sonyc.

\section{Conclusions}
We have carried out a near infrared spectroscopic analysis of 30 objects in the UpSco star forming region, 24 of which had been previously 
identified as brown dwarf candidates based on their photometry and proper motion alone.   
The resulting spectra confirm that all 24 are young very low mass objects with spectral types that range from M5.5 to L1.1.   

We have observed a diversity in form among the spectra that can impact on spectral type determination.   
Class II objects display excess emission and veiling in their spectra.   
The same form of excess emission and veiling is also present in the spectra of 7 out of 18 ($\sim 40$\%) class III objects.   
This is evidence that these class III objects are still accreting from an inner disc of dust and gas which is too faint to be detected via 
mid-infrared photometry.   

We present a catalogue of near infrared spectra for young brown dwarfs that are clear from discernible contributions from anything other than 
their photospheres.   We recommend the use of these objects as spectroscopic templates for identifying young brown dwarfs, and make them available 
at: http://browndwarfs.org/sonyc.   

\section*{Acknowledgements}

This work was supported by Science Foundation Ireland
within the Research Frontiers Programme under grant no. 10/RFP/AST2780.
This publication makes use of data products from the Wide-field
Infrared Survey Explorer, which is a joint project of the University of California, Los Angeles,
and the Jet Propulsion Laboratory/California Institute of Technology, funded by 
the National Aeronautics and Space Administration.
This publication also makes use of data products from the Two Micron 
All Sky Survey, which is a joint project of the University of Massachusetts and the 
Infrared Processing and Analysis Center/California Institute of Technology, funded by 
the National Aeronautics and Space Administration and the National Science Foundation.
DP would like to acknowledge that her contribution was partially supported by a grant from the American Astronomical Society.   
We would also like to thank the UKIDSS Team for the excellent database they have made 
available to the community.

\begin{table*}
 \begin{minipage}{170mm}
  \caption{UKIDSS Z photometry, Class, H peak index \& Spectral Type, H$_{2}$O index \& Spectral Type, H$_{2}$O-K2 index \& Spectral Type 
of the 30 objects observed with SpeX.   Objects are listed in the same order as in Table 1.}
  \begin{tabular}{|c|c|c|c|c|c|c|c|c|c}
  \hline
    Object & 2MASS Name & Z Mag. & Object & H peak & H peak & H$_{2}$O & H$_{2}$O & H$_{2}$O-K2 & H$_{2}$O-K2\\
 Number & & & Class & index & Sp. Type & index & Sp. Type & index & Sp. Type\\
 \hline
1 & 2MASSJ15420830-2621138 & 15.15 & III & 1.05 & M7.2 & 1.01 & M5.8 & 0.80 & M5.7 \\
2 & 2MASSJ15433947-2535549 & 18.15 & II & 1.36 & M9.5 & 1.16 & M9.4 & 0.61 & L0.1 \\
3 & 2MASSJ15442275-2136092 & 16.55 & III & 0.92 & M6.2 & 1.01 & M5.9 & 0.77 & M6.3 \\
4 & 2MASSJ15465432-2556520 & 14.16 & II & 1.06 & M7.3 & 1.01 & M5.9 & 0.80 & M5.7 \\
5 & 2MASSJ15472572-2609185 & 15.40 & II & 1.14 & M7.9 & 1.04 & M6.6 & 0.79 & M5.9 \\
6 & 2MASSJ15490803-2839550 & 14.82 & III & 0.97 & M6.6 & 0.99 & M5.5 & 0.81 & M5.3 \\
7 & 2MASSJ15491602-2547146 & 14.31 & III & 0.99 & M6.7 & 1.01 & M5.9 & 0.79 & M5.9 \\
8 & 2MASSJ15492909-2815384 & 14.29 & III & 1.07 & M7.3 & 1.03 & M6.4 & 0.77 & M6.5 \\
9 & 2MASSJ15493660-2815141 & 14.66 & III & 1.06 & M7.3 & 1.03 & M6.3 & 0.79 & M6.0 \\
10 & 2MASSJ15501958-2805237 & 16.04 & II & 1.07 & M7.4 & 1.03 & M6.4 & 0.76 & M6.7 \\
11 & 2MASSJ15514709-2113234 & 15.12 & II & 1.13 & M7.8 & 1.04 & M6.6 & 0.90 & M3.3 \\
12 & 2MASSJ15521088-2125372 & 15.72 & II & 1.21 & M8.5 & 1.07 & M7.2 & 0.88 & M3.7 \\
13 & 2MASSJ15524857-2621453 & 14.62 & III & 1.04 & M7.1 & 1.02 & M6.1 & 0.79 & M6.0 \\
14 & 2MASSJ15544486-2843078 & 15.51 & III & 1.04 & M7.1 & 1.01 & M5.9 & 0.78 & M6.1 \\
15 & 2MASSJ15551960-2751207 & 15.60 & III & 1.13 & M7.8 & 1.04 & M6.6 & 0.73 & M7.4 \\
16 & 2MASSJ15572692-2715094 & 14.93 & III & 0.94 & M6.4 & 1.00 & M5.5 & 0.84 & M4.7 \\
17 & 2MASSJ15582376-2721435 & 14.35 & III & 1.00 & M6.9 & 1.00 & M5.7 & 0.80 & M5.6 \\
18 & 2MASSJ15591513-2840411 & 14.14 & III & 1.00 & M6.8 & 0.99 & M5.5 & 0.80 & M5.7 \\
19 & 2MASSJ16002535-2644060 & 14.38 & III & 1.05 & M7.2 & 1.01 & M5.9 & 0.78 & M6.2 \\
20 & 2MASSJ16005265-2812087 & 15.04 & III & 1.05 & M7.2 & 1.02 & M6.1 & 0.75 & M6.8 \\
21 & 2MASSJ16062870-2856580 & 14.90 & III & 1.09 & M7.5 & 1.02 & M6.2 & 0.82 & M5.2 \\
22 & 2MASSJ16090168-2740521 & 14.33 & III & 1.13 & M7.8 & 1.03 & M6.2 & 0.75 & M6.9 \\
23 & 2MASSJ16101316-2856308 & 15.67 & III & 1.12 & M7.8 & 1.09 & M7.7 & 0.70 & M8.0 \\
24 & 2MASSJ16195827-2832276 & 18.74 & III & 1.45 & L0.3 & 1.24 & L1.1 & 0.59 & L0.8 \\
\\
25 & 2MASSJ15502934-2835535 & 16.05 & III & 0.99 & M6.7 & 1.03 & M6.4 & 0.75 & M6.9 \\
26 & 2MASSJ15504920-2900030 & 14.35 & III & 0.91 & M6.2 & 0.98 & M5.2 & 0.90 & M3.4 \\
27 & 2MASSJ15551768-2856579 & 14.32 & III & 0.96 & M6.5 & 0.99 & M5.4 & 0.93 & M2.6 \\
28 & 2MASSJ16035601-2743335 & 14.41 & III & 0.91 & M6.1 & 0.98 & M5.3 & 0.88 & M3.7 \\
29 & 2MASSJ16130482-2711214 & 15.56 & III & 0.92 & M6.2 & 0.99 & M5.4 & 0.89 & M3.5 \\
30 & 2MASSJ16190983-2831390 & 16.63 & III & 1.20 & M8.4 & 1.13 & M8.7 & 0.72 & M7.5 \\

\end{tabular}
\end{minipage}
\end{table*}

\begin{table*}
 \begin{minipage}{170mm}
  \caption{UKIDSS Z photometry, HPI Spectral Type, H$_{2}$O Spectral Type, H$_{2}$O-K2 Spectral Type and T$_{eff}$ 
of the 11 objects selected as near infrared spectral templates for young low mass stars and brown dwarfs between M5 and L1.}
  \begin{tabular}{|c|c|c|c|c|c|c}
  \hline
    Object & 2MASS Name & Z Mag. & HPI & H$_{2}$O & H$_{2}$O-K2 & T$_{eff}$ (K)\\
 Number & & & Sp. Type & Sp. Type & Sp. Type \\
 \hline

18 & 2MASSJ15591513-2840411 & 14.14 & M6.8 & M5.5 & M5.7 & 3,050\\
17 & 2MASSJ15582376-2721435 & 14.35 & M6.9 & M5.7 & M5.6 & 3,050\\
14 & 2MASSJ15544486-2843078 & 15.51 & M7.1 & M5.9 & M6.1 & 3,000\\
19 & 2MASSJ16002535-2644060 & 14.38 & M7.2 & M5.9 & M6.2 & 3,000\\
13 & 2MASSJ15524857-2621453 & 14.62 & M7.1 & M6.1 & M6.0 & 3,000\\
20 & 2MASSJ16005265-2812087 & 15.04 & M7.2 & M6.1 & M6.8 & 2,950\\
21 & 2MASSJ16062870-2856580 & 14.90 & M7.5 & M6.2 & M5.2 & 3,050\\
9 & 2MASSJ15493660-2815141 & 14.66 & M7.3 & M6.3 & M6.0 & 2,950\\
8 & 2MASSJ15492909-2815384 & 14.29 & M7.3 & M6.4 & M6.5 & 2,950\\
23 & 2MASSJ16101316-2856308 & 15.67 & M7.8 & M7.7 & M8.0 & 2,750\\
24 & 2MASSJ16195827-2832276 & 18.74 & L0.3 & L1.1 & L0.8 & 2,300\\

\end{tabular}
\end{minipage}
\end{table*}

\bsp

\label{lastpage}

\end{document}